\def\PRB{Phys. Rev. B, }
\def\PL{Phys. Lett., }
\def\PRD{Phys. Rev. D, }
\def\JPC{J. Phys. C, }

\def\PRL{Phys. Rev. Lett., }

\def\Tr{\rm Tr}
\def\ltwid{\mathrel{\raise.3ex\hbox{$<$\kern-.75em\lower1ex\hbox{$\sim$}}}}

\documentstyle[preprint,aps]{revtex}
\begin{document}
\draft
\title{Charge density oscillations in a quasi-two-dimensional electron\\
 gas for integer filling factors  }
\author {S. Sakhi and P. Vasilopoulos}
\address{ Department of Physics, Concordia University
1455 de Maisonneuve Blvd O
Montr\'eal, Qu\'ebec, Canada H3G 1M8
}
\maketitle

\begin{abstract}

The possibility of charge density oscillations in a {\it finite-thickness}
two-dimensional system is investigated for strong magnetic fields and
integer filling factors. Using an effective action formalism, it is shown
that an {\it oscillatory charge density} (OCD) is generated in a
self-consistent way
and is favored energetically over homogeneous  distributions.
It is smooth on the scale of the sample thickness and of the magnetic
length. The modulus of its wave vector is shown to be
experimentally accessible. The Hall voltage and the current density are
shown to {\it oscillate} with the same wave vector when a weak current is
applied. The stability of the
charge oscillations against impurity potentials is discussed.
\end{abstract}

Two-dimensional electron gas systems (2DEG) subject to a
magnetic field normal to the plane of confinement exhibit a rich
phase structure. This includes the manifestation of the integer and fractional
quantum Hall
effect \cite{PraGir} at low temperature and the expected Wigner crystallization
at
low densities \cite{Lau}.

In this letter, we study a system of interacting nonrelativistic charged
fermions in 3+1
dimensions embedded in a uniform neutralizing background of opposite charge.
This system is subjected to a uniform
perpendicular magnetic field  and the particles interact with each other
through
the repulsive Coulomb interaction.
The goal of the study is to analyze the low energy effective theory describing
this system and to explicitly show the possibility of charge density
oscillations.

{}From the study of strictly
two-dimensional systems in the absence of magnetic fields we know that the
ground state is a fluid of constant density \cite{Janc}. Furthermore,
it has been realized that a magnetic field driven charge density redistribution
in
the bulk is possible as
long as no Landau level is partially occupied \cite{MacDonald,Thouless}.
However, in these studies, the self-consistent equations used to describe the
charge density do not allow for oscillatory type solutions. Here, we show
that these are permissible and favored energetically over the uniform
distribution for real systems with {\it finite-thickness} when only the lowest
subband is occupied.

The approach we use relies on the effective action formalism combined with a
derivative expansion technique \cite{AitFra,Sakhi}. Here, the effective action
is simply defined
as the sum of all connected one-particle irreducible vacuum diagrams in the
presence of any background field \cite{Col}. This method has been extremely
useful, among other things, in showing the origin of the Chern-Simons term in
2+1
dimensions \cite{BabDasPan} and in analyzing the phase diagram of planar
superconductivity \cite{MacPanSak}. Here, after removing the four-Fermi
Coulomb term in favour of a Yukawa-type coupling to an auxiliary field
$\phi$, the effective action is obtained by integrating out the fermions.
The background
fields in the expansion take account of the magnetic field which gives rise to
Landau levels as underlying single particle states and also of a
self-consistent electrostatic potential. The result of the latter is to shift
the
center of the Landau wavefunctions and thereby make the electron density
position dependent. In this
way it is possible to maintain an integer filling factor throughout the
sample despite charge redistribution. Most importantly, we show explicitly
that an {\it oscillatory charge density} (OCD) can be generated in a
self-consistent
way as long as each Landau level is fully occupied. It is smooth on the scale
of the
magnetic length $\ell_B$ and of the thickness $d$. The modulus $q$ of its wave
vector, in the plane of the 2DEG, is given by
\begin{equation}
q\approx q_w-\frac{1}{\nu}\frac{\epsilon_0\hbar\omega_c}{e^2}
\end{equation}
Here $q\ell_B\ll 1$, $qd_w\ll 1$,
$\omega_c=|e|B/m^*  $ is the cyclotron frequency, $m^*$ the effective mass,
and $\epsilon_0$ the static dielectric constant which is assumed spatially
homogeneous. The Hall potential and the current density both
oscillate with the same wave vector. A possible consequence of this oscillatory
charge
density is the destruction of incompressibility in line with experimental and
theoretical studies of finite-thickness effects in a different context
\cite{Suen,Zhang}.

The action $S$, at finite temperature $T$, describing the interacting {\it
finite-thickness} 2DEG is given by
\begin{equation}
\label{action}
S=\int_0^{\beta}d\tau \int d^3x
\psi^\dagger \left(\partial_\tau +\mu -H_0\right)
\psi\nonumber
-\frac{e^2}{2\epsilon_0}\int_0^{\beta}d\tau \int \frac{d^3x d^3y}
{|{\bf x-y}|}\left[\psi^\dagger\psi({\bf
x},\tau)-\bar{\rho}\right]\nonumber\left[\psi^\dagger\psi({\bf y},\tau)
-\bar{\rho}\right];
\end{equation}
the first term is the kinetic energy with a parabolic dispersion
relation for the fermions; $H_0=(-i\hbar\nabla+e{\bf
A})^2/2m^*+U(z)$, the vector potential is ${\bf A}=(-By,0,0)$, $\beta =1/T$
($k_B=1$) and $\mu$ is the chemical potential. The potential $U(z)$
describes the confinement of the 2DEG in the $z$ direction. It can be that
of a square, triangular or parabolic well. Finally, the last term
describes the Coulomb interaction between the charged fermions.

In order to make $S$ bilinear in
terms of the fermion fields, we add to it a term of the form
\begin{equation}
\delta S=-\frac{\epsilon_0}{8\pi}\int d\tau\int d^3{\bf x}\left[\nabla\phi({\bf
x},\tau)\right]^2
\end{equation}
and perform the shift
$
\phi({\bf x},\tau)\longrightarrow\phi({\bf x},\tau)+(ie/\epsilon_0)
\int (d^3{\bf y}/|{\bf x -y}|)
\psi^\dagger\psi({\bf y},\tau),$
which cancels the Coulomb interaction term in (2). The
partition function is obtained by integrating
over the fermion field $\psi$ as well as over the scalar field $\phi$
. Clearly, this procedure has no effect since the integration
over $\phi$ gives a physically irrelevant constant; however, it
allows us to cancel the quartic term and thereby make $S$ bilinear
in the fermion fields. The linearized action reads
\begin{equation}
S=\int_0^\beta d\tau\int\,d^3x\left\{\psi^\dagger\left(\partial_\tau +\mu
-H_0-ie\phi({\bf x},\tau)\right)\psi
-\frac{\epsilon_0}{ 8\pi}\left[\nabla\phi({\bf x},\tau)
\right]^2+ie{\bar\rho}\phi({\bf x},\tau)\right\}
\end{equation}
It is important to emphasize that  the scalar
field $\phi$ in Eq. (4) couples to the charge density
$\psi^\dagger\psi$; its fluctuations in the presence of the uniform
magnetic field will describe the magnetoplasmon collective excitations.

The advantage of the action (4) is that its bilinear nature in the
fermion fields allows one to eliminate them exactly. The result of this
operation is the normalized effective action $S_{\rm eff}$ given by
\begin{eqnarray}
S_{\rm eff}&=&\int d\tau d^3{\bf x}\left\{\frac{\epsilon_0}{8\pi}
\left[\nabla\phi({\bf x},\tau)\right]^2-ie{\bar\rho}\phi({\bf
x},\tau)\right\}\nonumber\\
&&-\Tr\ln (\partial_\tau+\mu-H_0-ie\phi)
(\partial_\tau+\mu-H_0-V)^{-1},
\end{eqnarray}
where $\Tr(\cdots)$ denotes the functional trace operation.  The potential
$V({\bf x})$ in Eq. (5) will be specified later.

In this formalism, a uniform charge distribution corresponds to taking
$\phi({\bf x},\tau)$ purely fluctuating. Here, we assume a nontrivial vacuum
structure corresponding to some average electric potential
$\phi_0({\bf x},\tau)=\phi_0({\bf x})$. This can be obtained by minimizing
the effective action  $\delta S_{\rm eff}/ \delta\phi({\bf x},\tau)=0$;
the result after the substitution $e\phi_0({\bf x})=-iV({\bf x})$, is
the usual Poisson equation
\begin{equation}
\nabla^2 V({\bf x})=\frac{4\pi e^2}{\epsilon_0}\left[{\bar\rho}-\langle {\bf
x},\tau|\frac{1}{\partial_\tau+\mu-H_0-V}|{\bf x},\tau\rangle\right]
=-\frac{4\pi e^2}{\epsilon_0}\delta\rho({\bf x}).
\end{equation}
Furthermore, the fluctuations around this configuration are described by
\begin{eqnarray}
S_{\rm eff}&=&-\frac{\beta\epsilon_0}{8\pi e^2}\int d^3{\bf x}\left[\nabla
V({\bf
x})\right]^2+\frac{\epsilon_0}{8\pi}\int d\tau d^3{\bf x}\left[\nabla\phi({\bf
x},\tau)\right]^2\nonumber\\
&&-\Tr\ln\left[1-\frac{ie}{\partial_\tau +\mu -H_0
-V({\bf x})}\phi({\bf x},\tau)\right].
\end{eqnarray}
The last term in $S_{\rm eff}$ can be evaluated by a derivative expansion
technique \cite{Sakhi}. In a diagrammatic language, it corresponds to
a sum of an infinite number of loops with external legs representing the
field $\phi$. The quadratic term in this expansion, defines an induced
propagator for the field $\phi$, and the higher order terms represent
self-interaction. Here we will restrict ourselves only to the
quadratic term giving
\begin{equation}
-\frac{e^2}{2\beta}\sum_{\omega}\int d^3{\bf x}d^3{\bf y}\phi({\bf
y},-\omega)\Pi({\bf x,y};\omega)\phi({\bf x},\omega),
\end{equation}
where $\Pi({\bf x,y};\omega)$ is the polarization tensor, represented by the
diagram in Fig. (1) and $\omega=2\pi n/\beta$ the Matsubara frequency. The
ground state
energy in the presence of the
nontrivial electrostatic potential is obtained by
integrating out the fluctuations $\phi$. In terms of the potential $V({\bf
x})$ and the
polarization tensor $\Pi({\bf x,y};\omega)$, the result is
\begin{equation}
{\cal E}_G=-\frac{\epsilon_0}{8\pi e^2}\int d^3{\bf x}\left[\nabla V({\bf
x})\right]^2
+\lim_{\beta\to \infty}\frac{1}{2\beta}\Tr\ln \left[1+\frac{4\pi
e^2}{\epsilon_0}\frac{1}{\nabla^2}\Pi({\bf x,y};\omega)\right].
\end{equation}
The evaluation of $\Pi({\bf x,y};\omega)$ requires the
knowledge of the single-particle states of the
Schrodinger operator $H_0+V({\bf x})$. Before we proceed into that we exploit
the fact that when a
small electric field is applied in the $x$-direction, a Hall voltage
develops in the
$y$-direction. As a consequence, the electrostatic potential that enters the
Schrodinger equation depends only on the $y$ coordinate. Then
the eigenfunctions of the Schrodinger operator take the form
$\exp(ik_xx)\psi_\ell(y-y_0)X_0(z)/\sqrt{L_x}$ with  $y_0=\ell_B^2k_x$
, $X_0(z)$ the lowest subband wavefunction in the $z$-direction and
$\psi_\ell(y)$ satisfying
\begin{equation}
\left(-\frac{{\hbar^2}}{2m^*}\frac{d^2}{dy^2}+\frac{1}{2}m^*\omega_c^2y^2+
V(y_0+y)-\varepsilon_{\ell,y_0}\right)\psi_\ell(y)=0;
\end{equation}
here, $\ell$ is the Landau-level index, and
$\varepsilon_{\ell,y_0}$ the eigenvalue measured relative to the lowest
subband energy. Provided that $V(y)$ is a slowly varying function
on the
scale of $\ell_B$, i. e., $V^{''}(y_0)\ll m\omega_c^2$, we can expand
$V(y+y_0)$ in a Taylor series
and replace it by $V(y_0)+yV^{'}(y_0)$. Then $\psi_\ell(y-y_0)=
\Phi_\ell(y-y_0^{'})$ and
\begin{equation}
\varepsilon_{\ell, k_x}=\hbar \omega_c\left(\ell+\frac{1}{2}\right)+V(y_0)
-\frac{1}{ 2m^*\omega_c^2}\left[V^{'}(y_0)\right]^2,
\end{equation}
where $ y_0^{'}=y_0-V^{'}(y_0)/m^*\omega_c^2$, and
$\Phi_\ell(y)$ is a harmonic oscillator function. The last two terms of
Eq. (11)
 can also be obtained by considering $V(y)$
as a perturbation and using non-degenerate perturbation theory.

As in Ref. \cite{MacDonald}  we assume that no Landau level is partially
occupied and that the initial charge distribution is homogeneous
in the 2D plane when $V(y)\equiv 0$. Then the charge density $\delta\rho(y,z)$
entering Poisson's equation is given by
\begin{equation}
\delta\rho(y,z)=\frac{e}{L_x}\sum_{\ell,k_x}
\left[\Phi_\ell^2(y-y_0^{'})-\Phi_\ell^2(y-y_0)\right]
X_0^2(z)
=\frac{e\nu}{2\pi\hbar\omega_c}V^{''}(y)X_0^2(z),
\end{equation}
where $\nu$ is the filling factor. The last equality follows from a Taylor
expansion and the assumption $\delta y_0
=2\pi\ell_B^2/L_x\ll \ell_B$; it can also be obtained if $\Phi_\ell(y-y_0^{'})$
is
evaluated by perturbation theory. In either case the entire Hilbert space
is used to obtain Eq. (12) as the perturbation $V(y)\approx V(y_0)+
yV'(y_0)$ connects a Landau
level with the higher and lower levels, e.g., $\Phi_0(y-y_0^{'})$ is
not orthogonal to $\Phi_1(y-y_0)$. This is in sharp contrast with
usual treatments, e.g., those of Refs. \cite{Fuk,Yosh}, which are restricted
to the subspace of the lowest Landau level wavefunctions.
Eq. (11) and the last term in
Eq. (12) cannot be obtained if we restrict ourselves to this subspace.
Notice that Eq. (12) can be written as
$\delta\rho(y,z)=\delta\rho(y)X_0^2(z)$ with $\delta\rho(y)=e\nu V^{''}(y)/2\pi
\hbar\omega_c$.

Alternatively, this redistribution of the electron density will create a
potential $V(y,z)$ given by
\begin{equation}
V(y,z)=-\frac{e}{\epsilon_0}\int dy'dz' \delta\rho(y',z')\ln\left[(y-y')^2+
(z-z')^2\right],
\end{equation}
where use has been made of the charge-neutrality condition
\begin{equation}
\int_{-L_y/2}^{L_y/2}dy\delta\rho(y)=\frac{e\nu}{2\pi\hbar\omega_c}
\left[V^{'}(L_y/2)-V^{'}(-L_y/2)\right]=0.
\end{equation}
Notice that Eq. (13) merely expresses the potential produced at a point
$(y,z)$ by a strip parallel to the $x$ axis, at point $(y^{'}, z^{'})$,
with charge density $\delta\rho(y^{'},z^{'})$.

We assume, in line with most experiments, that only the lowest subband in the
$z$-direction is occupied. The electrons can then be
thought of as extended rod-like charges in the $z$-direction which are
allowed to move only in the $xy$ plane. For these reasons, the potential that
enters the Schrodinger equation is approximated by
the average $V(y)\approx <X_0(z)|V(y,z)|X_0(z)>$.
Notice that when the thickness of the 2DEG is zero
, the integral over $z$ is easily carried out and yields
Eq. (12) of Ref. \cite{MacDonald}.

Differentiating Eq. (13) twice with respect to $y$ and using Eq.
(12), gives the following integral equation
\begin{equation}
\delta\rho(y)=-\frac{e^2\nu}{\epsilon_0 \pi\hbar\omega_c}
\int X_0^2(z) dz\int X_0^2(z')  dz'
\int_{-L_y/2}^{L_y/2}dy'\delta\rho(y')
\frac{(z'-z)^2-(y'-y)^2}{\left[(z'-z)^2+(y'-y)^2\right]^2}.
\end{equation}
An oscillatory charge density $\delta\rho
(y)=\delta\rho(q)\sin(qy)$, with $q>0$ is a solution. When substituted on both
sides of Eq. (15), it gives the dispersion relation that $q$ has to satisfy
\begin{equation}
\eta=\int dz X_0^4(z)
-\frac{q}{2}\int X_0^2(z)
dz\int X_0^2(z')  dz'\, e^{-q|z'-z|},
\end{equation}
where $\eta=\epsilon_0\hbar\omega_c/2e^2\nu$. An oscillatory charge density of
the form $\delta\rho(y)=\delta\rho(q)\cos(qy)$
is another solution to this integral equation; however, it is not physically
permissible since it violates the charge neutrality condition (14) which
allows only odd solutions.

We evaluate this dispersion for a square well
$X_0(z)=
\sqrt{2/d}\cos(\pi z/d)$, $-d/2\leq z\leq d/2$. With $s=qd$, we obtain
\begin{equation}
\eta d=\frac{2\pi^2}
{s^2+4\pi^2}+\frac{16\pi^4}{s(s^2+4\pi^2)^2}(1-e^{-s}).
\end{equation}
When $qd\ll 1$, this gives Eq. (1)
with $q_w=3/d$.
For a triangular well we use $X_0(z)=(b_0^{3/2}/
\sqrt{2})z \exp(-b_0z/2)$ with $0\leq z,z^{'}\leq \infty$. For the average
thickness $d_0=3/b_0$ we obtain, for $qd_0\ll 1$,
again Eq. (1) with $q_w=9/8d_0$. Eq. (1) is also valid
for a parabolic well of frequency
$\omega_z$, for which $X_0(z)=(1/\pi\ell_z^2)^{1/4}
\exp(-z^2/2\ell_z^2)$, with  $q_w=\sqrt{2\pi/\ell_z}$.

This solution for $\delta\rho (y)$ and Eq. (12) give
$V(y)=-(2\pi\hbar\omega_c/e\nu q^2) \delta\rho (q) \sin(qy).$
For completeness, we also give the Hall current density,
averaged along the $z$-direction. By noting that ${\dot x}=\omega_c
(y-y_0)$, it is given by
\begin{equation}
J_x(y)=-\frac{e\omega_c}{L_x}\sum_{\ell,k_x}(y-y_0)\Phi_\ell^2(y-y_0^{'})
=\frac{e\nu}{2\pi\hbar}V'(y)
\end{equation}
where the same approximations have been made as in the derivation of
Eq. (12) {\it i.e.} $V\ll \hbar\omega_c$ and
$[V'(y_0)]^2/m\omega_c^2\ll \hbar\omega_c$. As was noted in Ref.
\cite{MacDonald}, Eq. (18) merely expresses the fact that the current
density is such that the Lorentz force is locally canceled by the
electrostatic potential. Furthermore, by integrating over $y$ we obtain
that the Hall current is proportional to the potential gradient, and the
constant of proportionality is $e^2\nu/ 2\pi\hbar$.

Now that the electrostatic potential profile is known we are able to evaluate
the ground state energy from Eq. (9). The
first term in this equation is the Hartree energy and is easily evaluated.
The result, per particle, is
\begin{equation}
w\equiv\frac{{\cal E}_G}{N_e}=-\frac{\pi^2\hbar^2}{\nu^2q^2m^*}\left[\delta
n(q)\right]^2,
\end{equation}
where $\delta n(q)=\delta \rho (q)/e$ and $N_e=\nu L_x L_y/2\pi\ell_B^2$
. The minus sign shows that the OCD state is
energetically favorable. This conclusion remains valid even when we include the
next term in the exact expression Eq. (9), here we give only the Fock
exchange energy contribution. This is obtained by keeping only the first
term when we expand the logarithm and pass to the zero
temperature limit which amounts to replacing $(1/\beta)\sum_\omega$ by
$(1/2\pi)\int d\omega$. The result is
\begin{equation}
\frac{{\cal E}^{\rm exc}_G}{N_e}=-\frac{e^2}{\epsilon_0 N_e}\int\frac{d^3{\bf
Q}}{(2\pi)^3}\frac{1}{{\bf Q}^2}\int d\omega \,\,\Im\Pi({\bf Q},\omega),
\end{equation}
which after some algebra becomes
\begin{equation}
\frac{{\cal E}^{\rm exc}_G}{N_e}=-\frac{e^2 L_x}{2\pi\epsilon_0
N_e}\sum_{\ell,\ell '=0}^{\nu -1}\int dk_x\int
dk_x'\int\int\frac{dQ_y dQ_z}{(2\pi)^2}\frac{F(Q_z) G_{\ell, \ell '}(u)}
{(k_x-k_x')^2+Q_y^2+Q_z^2}.
\end{equation}
Here
\begin{eqnarray}
u&=&\frac{Q_y^2\ell_B^2}{2}+\frac{\ell_B^2}{2}\left[k_x'-k_x+\frac{V'(\ell_B^2
k_x)-V'(\ell_B^2 k_x')}{\omega_c}\right]^2,\nonumber\\
G_{\ell, \ell '}(u)&=&\bigl(\ell!\,/\ell^\prime!\bigr)e^{-u}\,u^{
\ell^\prime-\ell}
\left[L_\ell^{\ell^\prime-\ell}(u)\right]^2,\nonumber\\
F(Q_z)&=&|\langle X_0|e^{izQ_z}|X_0\rangle|^2,
\end{eqnarray}
and $L_\ell^{\ell^\prime-\ell}$ is an associated Laguerre polynomial for
$\ell^\prime\geq\ell $ (for  $\ell^\prime\leq\ell $, interchange $\ell$ and
 $\ell^\prime$ in this formula).
This Fock exchange energy is a very complicated functional of the
potential $V$ or equivalently of the charge density amplitude
$\delta\rho(q)$. It can be verified after a straightforward algebra
that the term linear in $\delta\rho(q)$ is zero. The
quadratic term is given, for $\nu=1$ and $d\ltwid\ell_B$
\begin{equation}
\frac{{\cal E}^{\rm exc}_G}{N_e}=-\left(0.615-0.051\frac{d}{\ell_B}\right)
\frac{e^2}{\epsilon_0\ell_B}
\left[\frac{\delta n(q)}{n}\right]^2
\end{equation}

So far, we have examined the OCD state at zero
temperature. It is not hard to extend the same analysis to include the
temperature effects. In order to do so, one has to evaluate the free energy
when the charge density oscillates. Denoting by $\delta F$
the difference in free energies between the normal, spatially homogeneous,
and the OCD, spatially inhomogeneous, states with fixed $N_e$. This
difference can be written in terms of the thermodynamic potential as
$\delta F=\Omega_{OCD}(\mu)-\Omega_0(\mu_0)+N_e(\mu-\mu_0)$. Because
$N_e=-\partial\Omega_0/\mu_0$,
$\delta F=\Omega_{OCD}(\mu)-\Omega_0(\mu)+1/2
(\Delta\mu)^2\partial^2\Omega_0/\partial\mu_0^2+\cdots$ where
$\Delta\mu=\mu-\mu_0$. Using $V(\theta)=V_0\sin\theta$, and
\begin{equation}
\frac{\Omega_{OCD}(\mu)}{N_e}=-T
\sum_{\ell=0}^\infty\int_0^{2\pi}\frac{d\theta}{
2\pi}\,\ln\left(1+e^{\frac{\mu-\epsilon_\ell(\theta)}{T}}\right),
\end{equation}
we obtain for low temperatures, and $|V_0|/T\gg 1$
\begin{equation}
\delta F=w-\frac{T{\sqrt T}}{{\sqrt{2\pi
|V_0|}}}e^{-\hbar\omega_c\gamma/T},
\end{equation}
where $\gamma=1/2-(\Delta\mu+|V_0|)/\hbar\omega_c$.
This gives a negative contribution, and hence proves the stability of the
OCD at low temperature.

The OCD remains also stable in the presence of weak disorder as is usually
the case in a high-mobility 2DEG. This conclusion is reached by computing the
impurity averaged Green's function
which satisfies  Dyson's equation. Now the relation defining the charge density
in Eq. (12) is modified by multiplying the squared wavefunctions
$\Phi_\ell(y)$ by the
appropriate spectral functions. A straightforward calculation reveals that
the only effect of disorder is Landau level broadening and Eq. (12) is
unaffected.

We now comment on the use of the approximation  $V(y)\approx
<X_0(z)|V(y,z)|X_0(z)>$. It has been used in the past to study the collapse
of the FQHE state in a quantum well\cite{Zhang}. It is reasonable as long as
the electron density is not too high and the quantum well is not too wide
such that the subbands are not close to each other, otherwise the wavefunction
$X_0(z)$ will be modified. Finally, in evaluating the ground state energy in
the
presence of the OCD we limited ourselves to the Hartree-Fock contribution
although Eq. (9) goes beyond this approximation. This justified by the fact
that the neglected terms come in higher powers of
$(e^2/\epsilon_0\ell_B)[\delta n(q)/n]$ and are small for strong
magnetic fields under the assumptions used to derive Eq. (12).

An estimate of the parameters that enter Eq. (1) shows
that the OCD is experimentally accessible. For instance, for a square well
based on GaAS with $d=140 \AA$, $m^*/m=0.07$, $\epsilon=12$ and $B=15$ Tesla we
obtain $\lambda=2\pi/q=2\mu m$. A consequence of the OCD is that the
Landau levels given by Eq. (11) oscillate with $y_0$ Fig. 1. This facilitates
interlevel transitions through the indirect gaps and may destroy the quantum
Hall effect.

In summary we have shown that for fully occupied Landau levels and strong
magnetic fields, a self-consistent oscillatory charge density (OCD) is
possible in a finite-thickness two-dimensional system, when the lowest
subband is occupied. The OCD, which is impossible for the ideal
two-dimensional system, is smooth on the scale of the magnetic length and
of the thickness. It is energetically favorable over the homogeneous
distribution at low temperature and remains stable against weak impurity
scattering.

\begin{figure}
\caption{ Polarization diagram (bubble) entering Eq. (8). Solid lines are
fermion propagators, while dashed lines are the scalar fields.
}
\end{figure}
\begin{figure}
\caption{ The first lowest Landau levels ($\ell=0$, $\ell=1$) as a function of
$y_0/\ell_B$, obtained from Eq. (11) when $V(y)=V_0\sin(qy)$ and
$q\ell_B=1/3$. The dashed and solid curves are for $V_0=0.1\hbar\omega_c$
and $V_0=0.45\hbar\omega_c$. The dotted curve shows the Fermi level $\mu$.
}
\end{figure}
\end{document}